# Chemical Complementarity between the Gas Phase of the Interstellar Medium and the Rocky Material of Our Planetary System

Haiyang Wang and Charles H. Lineweaver

*Planetary Science Institute, Research School of Astronomy and Astrophysics, Research School of Earth Sciences, The Australian National University, Canberra, ACT, 2611, Australia.*

**Summary:** We compare the elemental depletions in the gas phase of the interstellar medium (ISM) with the elemental depletions in the rocky material of our Solar System. Our analysis finds a high degree of chemical complementarity: elements depleted in the gas phase of the ISM are enriched in the rocky material of our Solar System, and vice versa. This chemical complementarity reveals the generic connections between interstellar dust and rocky planetary material. We use an inheritance model to explain the formation of primordial grains in the solar nebula. The primary dust grains inherited from the ISM, in combination with the secondary ones condensed from the solar nebula, constitute the primordial rocky material of our planetary system, from which terrestrial planets are formed through the effects of the progressive accretion and sublimation. The semi-major-axis-dependence of the chemical composition of rocky planetary material is also observed by comparing elemental depletions in the Earth, CI chondrites and other types of carbonaceous chondrites.

## Introduction

Investigations of UV spectra of stars since the 1970s have revealed interstellar absorption features produced by atoms in their favoured ionization stages in the interstellar medium (ISM) [1]. Atomic abundances of heavy elements relative to that of hydrogen are below the reference cosmic abundances that implicitly refer to solar abundances. The reduction of heavy elements represents the missing atoms in the ISM gas phases. This feature is reinforced by the correlation of the increase of elemental depletions of the ISM gas phases with increasing condensation temperature, which is a proxy for how refractory an element is. In a study of abundances along 243 different sight lines from more than 100 papers, Jenkins (2009) [1] characterized the systematic patterns for the depletions of 17 different elements (C, N, O, Mg, Si, P, S, Cl, Ti, Cr, Mn, Fe, Ni, Cu, Zn, Ge, and Kr), from which he constructed a unified quantitative scheme that could lead to a better estimate of dust compositions.

The rocky material of our Solar System may be representative of rocky material in other planetary systems. The depletion of volatile elements relative to the solar abundance is a characteristic signature in primitive meteorites [2][3][4] and terrestrial planets [5][6][7][8]. The search for 'mysteries' (volatile-rich component) [9] continues with the search for "lost planets" [10], a scenario that is similar to the search for "missing atoms" in the ISM gas phase from Field (1974) [11] to Jenkins (2013) [12]. Lewis et al. (1987) [13] reported the discovery of presolar grains in carbonaceous chondrites. The major types of presolar grains like diamond, silicon carbide, and graphite found in meteorites are good noble gas carriers. Their abundances are quite diagnostic of nebular and parent-body thermal events, since these different presolar grains are destroyed at different temperatures [4]. Extensive investigations of carbonaceous chondrites [14][15] have shown a correlation between presolar grain abundances and bulk meteorite elemental abundance patterns. Based on the examination of interstellar gas and dust composition and comparison with meteoritic data, Yin (2005) [16] revealed that the depletion patterns of moderately volatile elements have a potential connection between the interstellar dust and the meteorites. The potential connection, or chemical complementarity, will be discussed further in this paper.

## Elemental Depletions in the ISM Gas Phase

Elements can be classified into three categories based on the range of the elemental condensation temperature ($T_C$): highly volatiles ($T_C < \sim 500$ K), moderately volatiles ($\sim 500$ K $< T_C < \sim 1400$ K), and refractories ($T_C > \sim 1400$ K), as shown in Fig. 1 for the simulated elemental depletions in the ISM gas phase. The x-axis is the 50% condensation temperature from Lodders (2003) [17] that indicates the volatility of an element. The y-axis is the logarithmic depletion factor (or the abundance ratio) of an element in the ISM gas phase relative to the cosmic reference abundances that are taken to be solar abundances. The abundance ratios of highly volatiles are approximately identical to the reference values while those of moderately volatiles are comparably lower than the reference values. The refractories are depleted considerably. The profound reductions of refractories and moderately volatiles inevitably prompt us to ask "where are these missing atoms?".

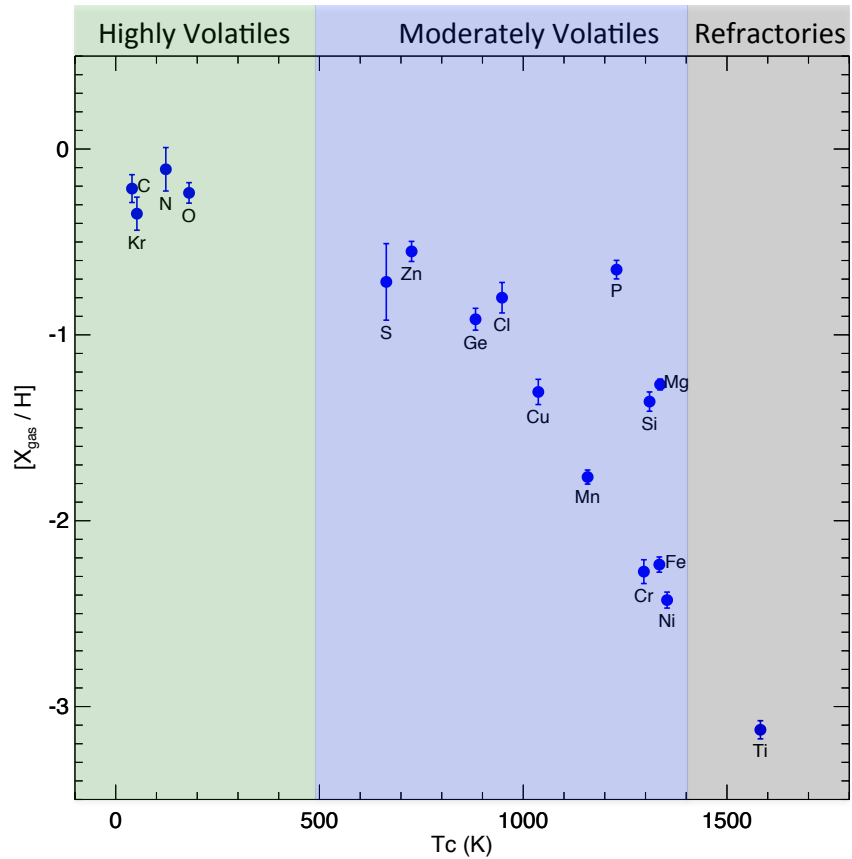

*Fig. 1: The logarithmic depletion factors of elements in the ISM gas phase as a function of the elemental condensation temperature ($T_C$) [17]. Elements are classified into three categories based on the range of $T_C$: highly volatiles ($T_C < \sim 500$ K), moderately volatiles ($\sim 500$ K $< T_C < \sim 1400$ K), and refractories ($T_C > \sim 1400$ K).*

The theoretical foundation for the above simulation is a unified quantitative scheme, developed by Jenkins (2009, 2013) [1][12], which relates the logarithmic depletion factor [$X_{gas}$/H] for an element $X$ to the sight-line parameter $F_*$ as shown in Eqn. 1. Fig. 2 illustrates the manner of parameterizing the depletion trends for different elements. The zero point $z_X$ for $F_*$ is chosen to make the covariance of the errors in the other two parameters, $A_X$ (slope) and $B_X$ (vertical offset) be zero.

$$[X_{gas}/H] = B_X + A_X(F_* - z_X) \quad (1)$$

The above unified form for elemental depletion in the gas phase is derived from a more straightforward parametric form Eqn. 2 that defines the logarithmic depletion of an element $X$ in terms of its depletion factor below the logarithm reference cosmic abundances ($\log(X/H)_{ref}$)

$$[X_{gas}/H] = \log[N(X)/N(H)]_{obs} - \log(X/H)_{ref} \quad (2)$$

where $N(X)$ is the column density of element $X$, and $N(H)$ represents the column density of hydrogen in both atomic and molecular form.

It is followed by the quantitative answer to the question "where are these missing atoms?". The missing atoms are presumably locked up in the form of dust grains, the complementary composition of which can be estimated from the above logarithmic depletion factor by

$$(X_{dust}/H) = (1-10^{[X/H]_{gas}}) \bullet (X/H)_{ref} \quad (3)$$

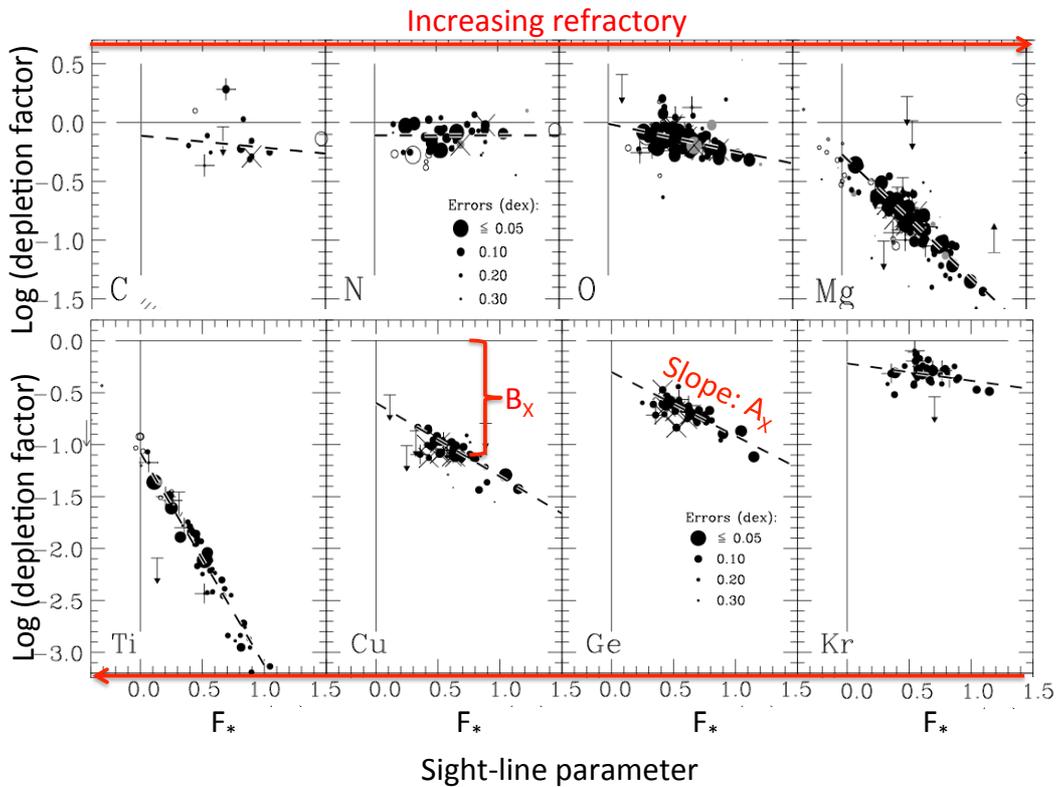

Fig. 2: Examples for 8 different elements that illustrate the manner of parameterizing the depletion trends in the ISM gas phase, adapted from [1][12]). The red arrows indicate the increasing refractory of elements in the upper and lower panels.

# Elemental Depletions in Rocky Material of Our Planetary System

Primitive meteorites and the Earth are the representatives of rocky material in our planetary system. Elemental abundances in CI chondrites, one of most primitive meteorites, are regarded as the best proxy for the solar abundances, with exceptions of the highly volatile elements H, C, N, O and the noble gases He, Ne, Ar, Kr, Xe [17][18]. Relative to CI chondrites, all other groups of meteorites, as well as the Earth, are considerably depleted in moderately volatile elements (see Fig. 3 and Fig. 4).

Fig. 3 is probably the most comprehensive comparison of elemental abundance patterns of different types of meteorites relative to CI chondrites plotted as a function of the condensation temperature, completed by Davis (2006) [4]. He described well the differences of abundance patterns of carbonaceous chondrites, ordinary chondrites and enstatite chondrites. Here, we would like to, nevertheless, emphasize their global similarities: (1) Moderately volatile elements are depleted, more or less, in all non-CI chondrites; (2) Refractory elements are approximately identical or slightly enriched compared to CI chondritic abundances, complementary to the depletion of their counterparts in the ISM gas phase as shown in Fig. 1; (3) The not-shown but well-accepted large depletion of highly volatiles in all chondrites compared with the proto-solar abundance is complementary as well to their ISM counterparts (Fig.1) that are approximately equal to the cosmic reference abundance.

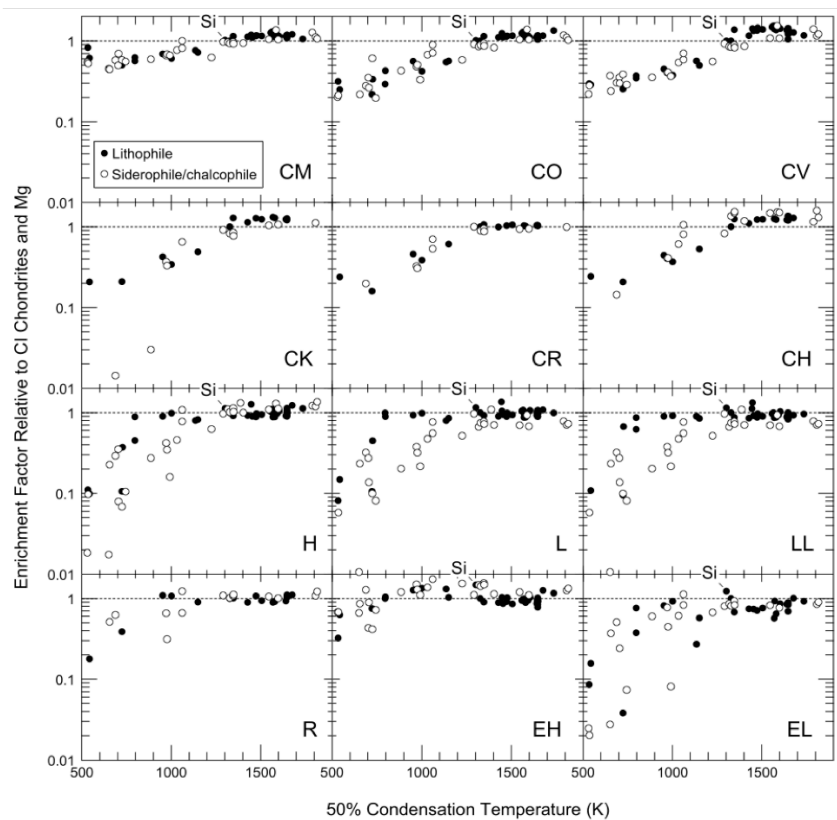

*Fig. 3: The comprehensive comparison of CI-normalized elemental abundance patterns of various primitive meteorites, inclusive of carbonaceous chondrites (CM, CO, CV, CK, CR, CH), ordinary chondrites (H, L, LL, R), and enstatite chondrites (EH, EL), plotted as a function of 50% condensation temperature [17]. The normalization-reference element is magnesium. The figure is from [4].*

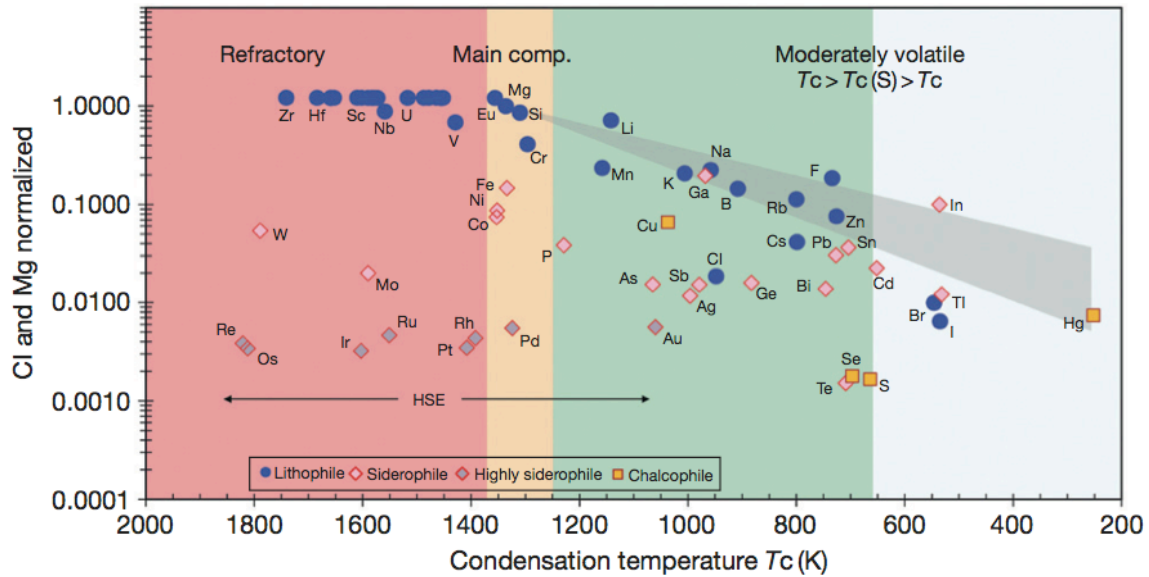

*Fig. 4: The comparison of CI-normalized abundances of geochemically classified elements in bulk silicate Earth plotted against the 50% condensation temperature [17]. The normalization-reference element is magnesium. The interesting elements for this cosmochemical analysis here are only lithophiles (indicated by blue dots). The figure is from [7].*

Fig. 4 shows the CI-normalized abundances of geochemically classified elements in the bulk silicate Earth (the bulk Earth excluding the core), plotted as a function of elemental condensation temperatures by Palme and O'Neil (2014) [7]. Here, we only care about lithophile elements (blue dots in Fig. 4), which are the dominant elements in bulk silicate Earth. It is clear that the abundance of refractory lithophiles (with condensation temperature greater than ∼ 1400 K) are approximately identical to that of CI chondrites, against the feature of the depletion of their counterparts in the ISM gas phase as shown in Fig. 1. The considerable depletion of moderately volatile elements is subject to further processes (e.g. evaporation) of the terrestrial building-blocks, consisting of the inherited pre-solar grains from the interstellar dust and the newly condensed solids from the solar nebula. The detailed interpretations are presented in the Discussion.

## Chemical Complementarity

From the above separate descriptions of elemental depletions in the ISM gas phase and in the inner Solar System rocky bodies, we see some potential complementary features between them. Yin (2005) [16] particularly discussed this topic from the comparison of dust composition and meteoritic data.

Three examples of interstellar data are plotted against the 50% condensation temperatures, as shown in Fig. 5a from [16]. As investigated in Fig. 1, refractory elements are highly depleted, and moderately volatile elements are progressively less depleted in the gas phase, indicating that these depleted elements are locked up in the dust. When compared to meteoritic data in Fig. 5b, the most intuitive and striking feature is that the abundances of interstellar gas phases are *complement*ary to the meteoritic data.

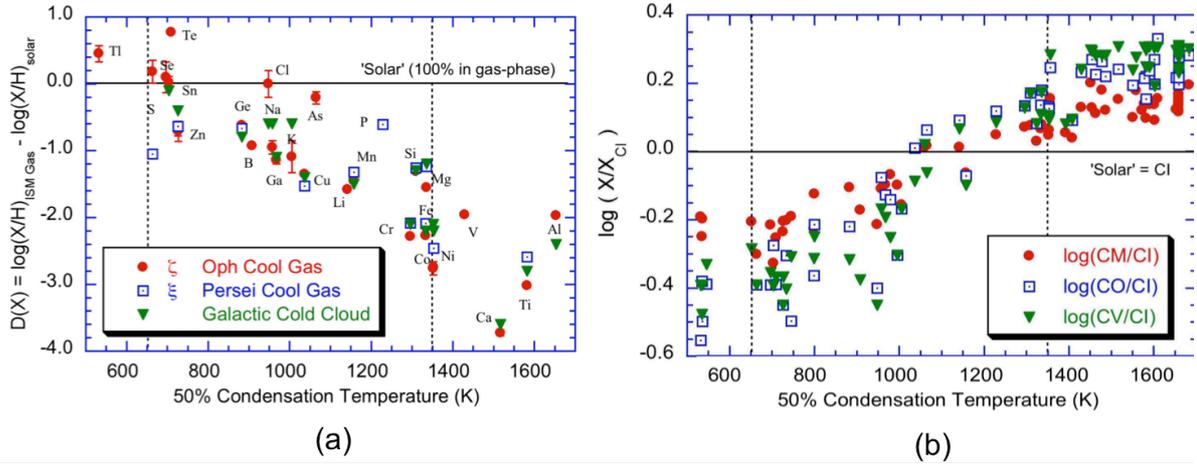

*Fig. 5: (a) Interstellar gas phase abundance of three examples relative to solar and (b) Primitive meteoritic abundance relative to CI, as a function of 50% condensation temperature [17]. For moderately volatile elements, interstellar gas phase data is 'mirror-imaged' by the meteoritic data. The figures are adapted from [16]*

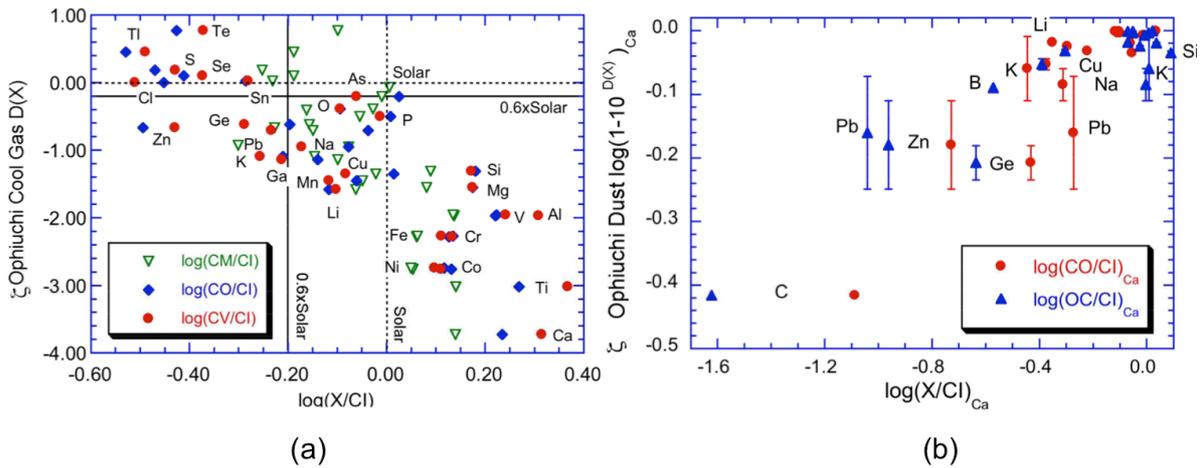

*Fig. 6: (a) The relative abundance of moderately volatile elements in the primitive meteorites is anti-correlated with that in interstellar gas. (b) The calculated ISM dust composition is positively correlated with that of primitive meteorites for moderately volatile elements. The figures are adapted from [16].*

Whereas, the depletion scales in the ISM gas phases and meteoritic data are apparently different by comparing the y-axes of Fig. 5a and 5b. Namely, this complementarity is not equivalent quantitatively. It is understandable. For interstellar dust, it is a long journey to form a 'primitive meteorite' after experiencing countless processes including shock-wave compression, condensation, sublimation, accretion, fractionation, burning (when passing through the Earth's atmosphere), and weathering (on the surface of the Earth). During these processes, its properties have been modified extensively. The majority of raw features in interstellar dust, therefore, are destroyed, and only a few tiny features are luckily reserved in forms of primitive meteorites that we can directly measure today. Among the few weakly reserved features, *chemical complementarity* between the ISM gas phases and the meteorites is the most striking one, whereas, the comparable scale between them is not expected.

Plotting the y-axis of Fig. 5a and Fig. 5b together as shown in Fig. 6a, it is clear that the interstellar gas is anti-correlated with the primitive meteoritic composition, underscoring that the provenance of the primitive meteorites is essentially the interstellar dust. In place of y-axis of Fig. 6a to the calculated dust composition from Eq. (3) as shown in Fig. 6b, the dust composition shows positive correlation with meteoritic composition for moderately volatile elements, further reinforcing the connections between primitive meteorites and the interstellar dust.

Based on the above comparisons, Yin (2005) [16] proposed an 'Inheritance Model' to explain the chemical connections between the interstellar dust and the material in the early active solar nebula. We schematically draw the mechanism in Fig. 7 to connect the diffuse interstellar medium to the protoplanetary disk, which corresponds to Yin (2005) [16]'s descriptions for these processes: (1) In the diffuse interstellar medium where elemental depletions are observed, the volatile elements are in the hot ionized gas phase, and the refractory elements are locked in the dust grains; (2) In the cold and dense molecular cloud stage, the gas phase consists of H and He only. Organic-rich icy mantles condense with all the volatile elements on to the refractory cores. Grain growth may have already occurred at this stage; (3) Collapse happens in the densest molecular cloud core with the continuous increase of surrounding gas pressure, leading to the rapid formation of the solar nebula with rotating gas and dust surrounding the central proto-star (namely the protoplanetary disk). The condensed layer of volatiles on most grain surfaces could easily be vaporized or sublimated when the grain is subjected to a reheating event, such as adiabatic compression or passage through shock waves prevalent during the collapse phase and in the early active solar nebula.

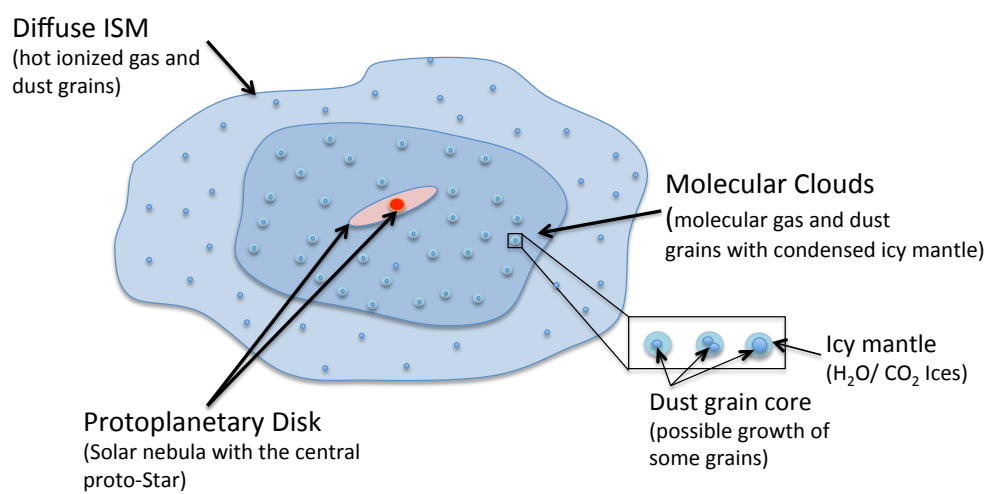

*Fig. 7: Schematic processes of dust grains inherited from the ISM to protoplanetary disk*

## Discussion

In addition to the chemical complementarity drawn from the comparison of elemental depletions between the ISM and the rocky planetary material, the other striking feature is the semi-major-axis dependency if investigating the elemental depletions in the inner solar system material specifically.

Fig. 8 is a comprehensive comparison of elemental depletion patterns among the bulk Earth, CI chondrites, and other non-CI carbonaceous chondrites (CCs) normalized to the proto-Sun by using the highly refractory aluminium as the normalization-reference element. Firstly, we can compare the normalized abundance of refractory (black dots), moderately volatile elements (green dots) in the non-CI CCs. From this comparison we can draw a similar conclusion as drawn from Fig. 3, namely the approximate equality of refractory elements and the increasing depletions of moderately elements, versus the decrease of the elemental condensation temperature, in the non-CI CCs relative to the Sun. Secondly and also more importantly, Fig. 8 additionally incorporates the elemental depletion pattern of the Earth ('blue' band in this figure) and that of CI chondrites ('red' band), from which we can clearly see their different elemental depletion trends. Contrary to the depletion trends in the ISM gas phases, the elemental depletion magnitudes in the rocky material of our planetary system are anti-correlated with the elemental condensation temperature. Meanwhile, considering the semi-major-axis of Earth (1 AU), CI (beyond 3 AU), and other CCs (within 2-3 AU), the depletion magnitude for the same element decreases with increasing distance to the Sun. The depletion slope and the depletion-transitional elements also vary correspondingly. Analysis of these patterns, various comparisons and implications will be investigated in detail in Wang et al. (2016) [8]. But a critical conclusion that we can learn from this comparison is that the elemental depletion in the rocky material of our planetary system is semi-major-axis dependent and dominated by the effect of evaporation or sublimation.

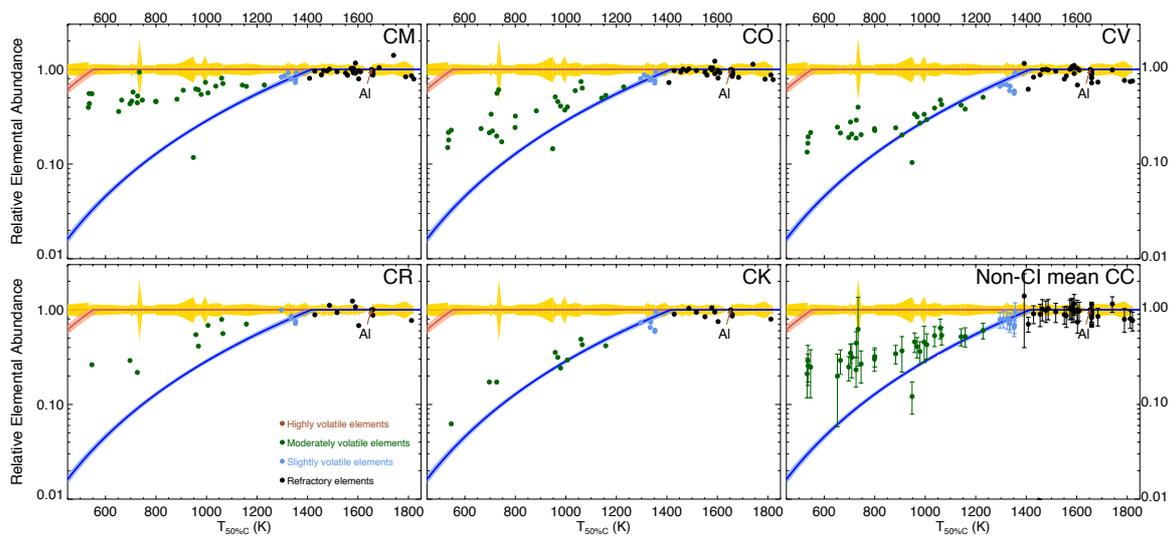

*Fig. 8: The comprehensive comparison of the elemental depletion patterns of the bulk Earth ('blue' band), CI chondrites ('red' band), and five types of non-CI carbonaceous chondrites along with their mean, relative to the proto-Sun. The normalization-reference element is aluminium. The feature of semi-major-axis-dependency of elemental depletions in the inner Solar System rocky material can be drawn from this comparison, analogous to the feature of chemical complementarity drawn from the comparison between the ISM and the inner Solar System rocky material.*

In addition, somebody may suspect whether this evaporation also works for giant gas planets like Jupiter. It deserves to be emphasized that evaporation or sublimation works through the Solar System, but it only has a significant influence on the rocky planets in the inner part of our system. Beyond the snowline, the effect of evaporation is considerably decreased, and enormous quantities of volatiles are able to survive and finally build up on those gas planets including Jupiter.

# Conclusions

Several conclusions can be derived from the analysis of elemental depletion in the gas phase of the ISM and that in the rocky material of the Solar System, in terms of the chemical evolution processes between them:

1. In the gas phase of the interstellar medium, refractory and moderately volatile elements are depleted. The missing atoms are locked in the interstellar dust or large molecules and are inherited by the material that forms a planetary system.

2. In the (terrestrial) planetary disk, depletions of moderately volatile elements show complementarity with the depletions in the ISM gas phase. The progressive processing of pre-solar grains inherited from the interstellar dust, along with newly formed solids in planetary disk, lead to planetesimals that form terrestrial planets.

3. The semi-major-axis-dependent elemental depletions in rocky planetary material are mainly subject to the effect of evaporation or sublimation, other than condensation that may govern the interstellar gas phase elemental depletion.

# Acknowledgment

We acknowledge that the idea of this work was inspired by Yin (2005), and relied on Davis (2006), Jenkins (2009, 2013), Palme and O'Neil (2014), and Wang et al. (2016). We also thank Michael Dopita and Ralph Sutherland for helpful comments.